\documentclass[aps,twocolumn,superscriptaddress,prl,amsmath,amssymb,10pt,aps,longbibliography]{revtex4-1} 
\usepackage[colorlinks=true,linkcolor=magenta,citecolor=magenta,urlcolor=magenta]{hyperref}
 \usepackage{lipsum}
 \usepackage{graphicx}
\usepackage{latexsym}
\usepackage{amsmath}
\usepackage{amsthm}
\usepackage{amssymb}
\usepackage{epstopdf} 
\usepackage{enumerate}
\usepackage{enumitem}
\usepackage{setspace}
\usepackage{dcolumn}
\usepackage{multirow}
\usepackage{bm}
\usepackage{mathrsfs}
\usepackage{setspace}
\usepackage{slashed}
\usepackage{color}
\usepackage{float}
\usepackage{youngtab}
\usepackage{tikz}
\usepackage{tikz-3dplot}
\usepackage{braket}
\usepackage{cancel}
\usepackage{subfigure}
\usepackage{pbox}
\usepackage{kotex}
\usepackage{array}
\usepackage{tabularx}
\usepackage[normalem]{ulem}

\usepackage{lipsum}
\graphicspath{{Figures/}}

\usetikzlibrary{shapes,snakes,arrows,chains,matrix,positioning,scopes,calc}
\usetikzlibrary{decorations.markings}

 \usepackage{moreverb}
\allowdisplaybreaks 
\begin{document}

\title{ Manipulating Topological Quantum Phase Transitions \\ of Kitaev's Quantum Spin Liquids with Electric Fields} 
      
\author{Pureum Noh}
	\thanks{These authors contributed equally: Pureum Noh, Kyusung Hwang.}
 \affiliation{Department of Physics, Korea Advanced Institute of Science and Technology (KAIST), Daejeon 34141, Korea}

\author{Kyusung Hwang}
	\thanks{These authors contributed equally: Pureum Noh, Kyusung Hwang.}
	\affiliation{School of Physics, Korea Institute for Advanced Study (KIAS), Seoul 02455, Korea}

 \author{Eun-Gook Moon}
\thanks{egmoon@kaist.ac.kr}
\affiliation{Department of Physics, Korea Advanced Institute of Science and Technology (KAIST), Daejeon 34141, Korea}


\begin{abstract} 
Highly entangled excitations such as Majorana fermions of Kitaev quantum spin liquids have been proposed to be utilized for future quantum science and technology, and a deeper understanding of such excitations has been strongly desired.
Here we demonstrate that Majorana fermion's mass and associated topological quantum phase transitions in the Kitaev quantum spin liquids may be manipulated by using electric fields in sharp contrast to the common belief that an insulator is inert under weak electric fields due to charge energy gaps. 
Using general symmetry analysis with perturbation and exact diagonalization, we uncover the universal phase diagrams with electric and magnetic fields.  
We also provide distinctive experimental signatures to identify Kitaev quantum spin liquids with electric fields, especially in connection with the candidate materials such as $\alpha$-RuCl$_3$.
 \end{abstract}

\maketitle
{\it Introduction: }  
Quantum spin liquids (QSLs) intrinsically host an enormous amount of quantum entanglement, which has attracted a great deal of interest in the research of future science and technology \cite{Zhou2017.4,Savary2016.11,Balents2010.3,Anderson1973.2,Knolle2019.3}. 
%
The intrinsic massive entanglement prevents quantum spin liquids from developing a trivial magnetic ordering, and instead emergent novel excitations may appear in QSLs.
%
Kitaev quantum spin liquid (KQSL) is one of QSL that has attracted significant attention \cite{Kitaev2005.10}. 
In KQSLs, the interactions between spin degrees of freedom are exactly solvable, leading to emergent Majorana fermions and Abelian or non-Abelian anyons.
These exotic properties make KQSLs promising platforms for topological quantum computation \cite{Nayak2008.9, Kitaev2002.5}.

The search for candidate materials that can exhibit KQSL has been a major challenge in the field of condensed matter physics.
In recent years, significant progress has been made in identifying and characterizing KQSL candidate materials, such as $\alpha$-RuCl$_3$ \cite{Plumb2014.7,Koitzsch2016.9,Sandilands2015.4,Kim2015.6,Kim2016.4,Winter2017.11,Winter2016.6,Winter2018.2,Yadav2016.11} and Na$_2$Co$_2$TeO$_6$ \cite{Takeda2022.11,Viciu2007.1,Imamura2023.5}, through various experiments \cite{Songvilay2020.12,Lin2021.9,Wulferding2020.3,Tanaka2022.1}. 
One of the unique features of KQSLs is their response to external magnetic fields, which can induce exotic phases such as a chiral spin liquid \cite{Kitaev2005.10}.
Despite the theoretical predictions, the experimental investigation of KQSLs in magnetic fields has remained challenging due to the need to explore a narrow range of magnetic field  \cite{Tanaka2022.1,Jiang2011.6,Janssen2016.12,Gohlke2018.7,Zhu2018.6,Hickey2019.1,Nasu2018.8,Liang2018.8,Yoshitake2020.3,Hwang2022.1,
Gordon2021.2,Ronquillo2019.4,Go2019.4,Vinkler2018.6,Ye2018.10}.
\begin{figure}[b]
\includegraphics[width=0.48\textwidth]{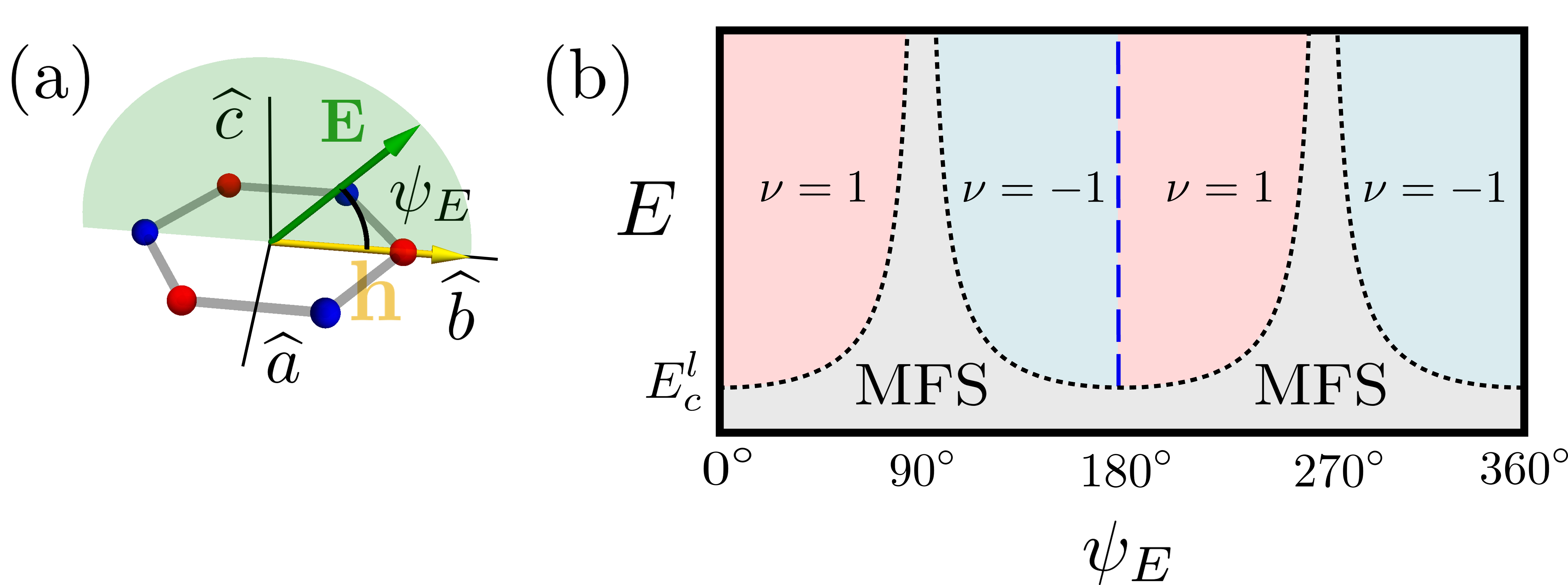}
  \caption{The universal topological phase transition by electric fields. (a) Diagram of the direction of electric ($\mathbf{E}$) and magnetic fields ($\mathbf{h}$). The magnetic field is in the direction of  $\hat{b}$, and the electric field lies in the $\hat{b}-\hat{c}$ plane. (b) The universal phase diagram at angles of the electric field ($\psi_E$) and the strength of the electric field ($E$) ranging from 0 to $|\mathbf{h}|$. The Chern number ($\nu$) is not defined due to the Majorana-Fermi surface (MFS) in the gray area, and the lower critical strength $E^{l}_c$ has the order of $\frac{|\mathbf{h}|^2}{\Delta_f}$. Dashed blue and dotted black lines indicate topological phase transitions.} 
 \label{fig1}
\end{figure}

In this letter, we demonstrate striking characteristics of electric-field-driven topological quantum phase transitions (TQPTs).
First, varying with the amplitude of electric fields ($E$), we find TQPTs between critical states and bulk energy-gapped states. The former (latter) states host a Fermi surface of Majorana fermions, also called Majorana-Fermi surface (MFS) \cite{Chari2021.4}, (topological invariants) for $E < E^{l}_c$ ($E > E^{l}_c$).
We remark that the presence of such TQPTs is in drastic contrast to the conventional wisdom in the literature that the size of the Fermi surfaces of Majorana fermions is proportional to $E$.
%
Second, by rotating an electric field, we find the possibility of the two types of TQPTs between the phases with opposite topological invariants.
One type is conventional in the sense that TQPTs appear with quantum critical points, but the other type permits in-between quantum critical states. 
Remarkably, the two types of TQPTs are only possible for the intermediate amplitude of electric fields because they are washed away for small enough electric fields and KQSLs become unstable for strong enough electric fields. 
By utilizing the characteristics, we also propose how to detect KQSLs in the candidate materials such as $\alpha$-RuCl$_3$.

\textit{Model Hamiltonian.} 
Let us consider the isotropic Kitaev model under electric ($\mathbf{E}$) and magnetic fields ($\mathbf{h}$) to be specific and discuss its generalization below. 
The Hamiltonian is   
%
%
\begin{align}
H(\mathbf{h}, \mathbf{E})=K \sum_{\langle i,j\rangle_{\gamma}}S^{\gamma}_iS^{\gamma}_j-\mathbf{h}\cdot \mathbf{S} -\mathbf{E}\cdot \mathbf{P},  \nonumber
\end{align}
where $\langle i,j \rangle_{\gamma} $ are for the nearest-neighbor bonds  with a component $\gamma \in \{x,y,z\}$ and $S^{\gamma}_j$ is  a $\gamma$ component spin operator at a site $j$ \cite{Kitaev2005.10}. 
The total spin operator is defined as $\mathbf{S}  = \sum_{j}\mathbf{S}_j$, and the interaction parameter $(K)$ for the bond-dependent exchange interaction is introduced. 

The explicit form of the electric polarization operator ($\mathbf{P}$) may be obtained by microscopic analysis \cite{Miyahara2016.1,Bolens2018.4,Bolens2018.9}, and for our purposes, it is enough to utilize the symmetry approach, following the previous works \cite{Bolens2018.9,Chari2021.4}. 
Since $\mathbf{P}$ is even under the time-reversal transformation and odd under the space-inversion transformation, the polarization operator becomes, 
$P^{\mu}\equiv \sum_{\langle i,j\rangle_{\gamma}}\mathbf{p}^{\mu}_{\gamma}\cdot (\mathbf{S}_i \times \mathbf{S}_j)$
 where $\mathbf{p}^{\mu}_\gamma$ is a vector with 27 components. 
 For the isotropic Kitaev model, only five of 27 parameters are independent \cite{Chari2021.4},  
and for clarity, we utilize the Hamiltonian,
\begin{align}
H(\mathbf{h}, \mathbf{E})= K \sum_{\langle i,j\rangle_{\gamma}}S^{\gamma}_iS^{\gamma}_j-\mathbf{h}\cdot \sum_j \mathbf{S}_j-\mathbf{E}\cdot\sum_{\langle i,j\rangle_{\gamma}} \mathbf{S}_i \times \mathbf{S}_j, \label{Hamiltonian} \nonumber
\end{align} 
in the main text as our prime example. We refer to supplementary materials (SM) for discussions about generic cases. 

Let us first consider the symmetries of the Hamiltonian. 
For an ideal monolayer system, the Hamiltonian $H(0, 0)$ enjoys $ \mathbb{D}_3$, where $ \mathbb{D}_3$ is for the dihedral group of order 6, in addition to the spacial inversion ($ \mathcal{P}$) and the time-reversal ($\mathcal{T}$). 
Turning on a magnetic field, the time-reversal and $ \mathbb{D}_3$ symmetries are completely broken except in certain directions of the magnetic fields. For example, the Hamiltonian $H(\mathbf{h} \parallel \hat{\mathbf{b}}, 0)$ only enjoys the two-fold rotational symmetry along $\hat{\mathbf{b}}$ and $\mathcal{P}$. See Figure \ref{fig1}(a) for the notation of the directions of fields. 

Physical quantities are characterized by representations of symmetry groups. For example, the thermal Hall conductivity ($\kappa_{ab}$) is odd under $\mathcal{T}$ and $C_2(\hat{\mathbf{b}})$, and it is even under $\mathcal{P}$ (Table \ref{table1}). 
It has been well understood that the two-fold rotational symmetry ($C_2(\hat{\mathbf{b}})$) protects the gapless condition of Majorana fermions in Kitaev quantum spin liquids. 

Turning on an electric field, all the symmetries are completely broken except in the two cases:
\begin{itemize}
\item $\mathbf{E} \parallel \hat{\mathbf{b}}$, \quad $\mathbb{G}_b \equiv \{ C_2(\hat{\mathbf{b}}), \big( C_2(\hat{\mathbf{b}}) \big)^2 \}$,
\item $\mathbf{E} \parallel \hat{\mathbf{c}}$, \quad $\mathbb{G}_c \equiv \{ \mathcal{P}C_2(\hat{\mathbf{b}}), \big( \mathcal{P}C_2(\hat{\mathbf{b}}) \big)^2 \}$, 
\end{itemize} 
where $\mathbb{G}_{b,c}$ are the symmetry groups for each case. We note that the case of $\mathbf{E} \parallel \hat{\mathbf{c}}$ is not invariant under $C_2(\hat{\mathbf{b}})$ and $\mathcal{P}$ symmetries but invariant under the combination, $\mathcal{P}C_2(\hat{\mathbf{b}})$. 
Below, we show that both $\mathbb{G}_b$ and $\mathbb{G}_c$ protect the gapless Majorana fermions in KQSLs though their effects have significant differences in terms of TQPTs.

\begin{table}[t]
\centering 
\begin{tabular}{>{\centering}p{3.5cm}||>{\centering}p{1.1cm}|>{\centering}p{1.1cm}|>{\centering}p{1.1cm}|>{\centering\arraybackslash} p{1.1cm}}
\hline 
Physical quantities & $\mathcal{T}$ & $\mathcal{P}$ & $C_2(\hat{\mathbf{b}})$ & $\mathcal{P}C_2(\hat{\mathbf{b}})$ \\ 
\hline \hline 
$\kappa_{ab}( \mathbf{h}, \mathbf{E})$ & odd & even & odd & odd \\
\hline
$\nu$ & odd & even & odd & odd \\
\hline
$m( \mathbf{h}, \mathbf{E})$ & odd & even & odd & odd \\
\hline
$h_x + h_y+h_z$  & odd & even & odd & odd \\ 
\hline
$h_x h_y h_z$ & odd & even & odd & odd \\
\hline
$E_c( E_a h_a + E_b h_b)$ & odd & even & odd & odd \\
\hline
$h_a (E_a^2- E_b^2) -  2  h_b E_a E_b$ & odd & even & odd & odd \\
\hline
\hline
$\mu( \mathbf{h}, \mathbf{E})$ & odd & odd & odd & even  \\
\hline 
$h_a E_b - h_b E_a$ & odd & odd & odd & even \\
\hline
$E_ch_b(h_b^2-3h_a^2)$ & odd & odd & odd & even  \\
\hline 
\end{tabular}

\caption{Symmetry properties of physical observables under the time-reversal ($\mathcal{T}$), inversion ($\mathcal{P}$) and the two-fold rotation ($C_2(\hat{\mathbf{b}})$). 
The thermal Hall coefficient ($\kappa_{ab}$), topological invariant ($\nu$), and the mass function $m( \mathbf{h}, \mathbf{E})$ are in the same representation while the chemical potential function ($\mu ( \mathbf{h}, \mathbf{E})$) is in a different representation. We also present the functions of electric and magnetic fields in the two representations. All the quantities are invariant under three-fold rotations. See SM for more detailed information. 
}
%
 \label{table1}
\end{table}

\textit{Weak electric and magnetic fields.} 
Following the original approach of Kitaev \cite{Kitaev2005.10}, we utilize perturbative calculations with the Majorana representation of quantum spins ($S^{\gamma}_j = i c_j b_j^{\gamma}$) with four Majorana fermions ($b_j^{\gamma}, c_j$) at a site $j$ for weak electric and magnetic fields ($|\mathbf{h}|, |\mathbf{E}| \ll K$). 
The low-energy effective Hamiltonian below the flux gap ($\Delta_f$) becomes 
\begin{align*}
&H_{{\rm eff}}(\mathbf{h}, \mathbf{E})=\frac{1}{2}\sum_{\mathbf{k}} 
\Psi_{\mathbf{k}}^{\dagger}\left(
\sum_{a=0,1,2, 3} \epsilon_a(\mathbf{k}, \mathbf{h}, \mathbf{E}) \tau^a 
\right)\Psi_{\mathbf{k}},
\end{align*}
with a two component spinor, $\Psi_{\mathbf{k}}=(c_{\mathbf{k},A}, c_{\mathbf{k},B})^{T}$, and $c_{\mathbf{r},A(B)}=\sqrt{\frac{2}{N}}\sum_{\mathbf{k}}e^{i\mathbf{k}\cdot \mathbf{r}}c_{\mathbf{k},A(B)} $. 
The identity and Pauli matrices in the sublattice spinor space  ($\tau^{0,1,2,3}$) are introduced with the energy functions, $\epsilon_{0,1,2,3}(\mathbf{k}, \mathbf{h}, \mathbf{E})$. 
The eigenenergy of the Hamiltonian is 
\begin{eqnarray}
E_{\pm}(\mathbf{k}, \mathbf{h}, \mathbf{E}) = \epsilon_{0}(\mathbf{k}, \mathbf{h}, \mathbf{E})\pm \sqrt{\sum_{a=1,2,3}\epsilon_{a}(\mathbf{k}, \mathbf{h}, \mathbf{E})^2 }. \nonumber
\end{eqnarray}
Without electric and magnetic fields, the energy functions vanish at the corners of the first Brillouin zone ($\mathbf{k} = \pm \mathbf{K}_M$), and the linear dispersion is determined by $\epsilon_1(\mathbf{k}, 0,0)$ and $\epsilon_2(\mathbf{k}, 0,0)$.
Thus, in the regime of weak electric and magnetic fields, the presence of energy-gap or Fermi-surfaces of Majorana fermions is mainly determined by the chemical potential function ($\mu( \mathbf{h}, \mathbf{E}) \equiv \epsilon_{0}( \mathbf{K}_M, \mathbf{h}, \mathbf{E})$)
and the mass function ($m( \mathbf{h}, \mathbf{E}) \equiv \epsilon_{3}( \mathbf{K}_M, \mathbf{h}, \mathbf{E})$). 
 If $|\mu( \mathbf{h}, \mathbf{E})| <|m( \mathbf{h}, \mathbf{E})|$, there is an energy gap, and the topological invariant ($\nu$) is given by the sign of $m( \mathbf{h}, \mathbf{E})$. As for the case of $|\mu( \mathbf{h}, \mathbf{E})| >|m( \mathbf{h}, \mathbf{E})|$, the topological invariant is not defined because of the presence of MFS.

One can understand the symmetry properties of the energy functions by extending the original discussion of the projective representation by Kitaev \cite{Kitaev2005.10} (see also SM). 
Note that $m( \mathbf{h}, \mathbf{E})$ is in the same representation of $\kappa_{ab}$ while $\mu( \mathbf{h}, \mathbf{E})$ is in a different representation since it is odd under the inversion symmetry.

Our strategy is to utilize the symmetry properties of  $m( \mathbf{E}, \mathbf{h})$ and  $\mu( \mathbf{E}, \mathbf{h})$, which can be applied beyond the pure Kitaev model. For simplicity, we consider a magnetic field along a bond direction and an electric field on the bc plane with an angle $\psi_E$, 
\begin{eqnarray} 
\mathbf{h} = h \hat{\mathbf{b}}, \quad \mathbf{E}=E (\cos{\psi_E}\hat{\mathbf{b}}+\sin{\psi_E}\hat{\mathbf{c}}). \nonumber 
\end{eqnarray}
We find that
\begin{eqnarray}
m( \mathbf{h}, \mathbf{E}) = c_m \frac{h E^2}{\Delta_f^2} \sin(2\psi_E), \quad \mu( \mathbf{h}, \mathbf{E}) &=& c_{\mu} \frac{h^3 E}{\Delta_f^3} \sin(\psi_E) \nonumber 
\end{eqnarray}
 up to the fourth order of electric and magnetic fields with two dimensionless constants ($c_m, c_{\mu}$). Note that the forms of $m( \mathbf{h}, \mathbf{E})$ and $\mu( \mathbf{h}, \mathbf{E})$ for generic field directions are presented in SM. 

A few remarks are as follows. 
First, the symmetry properties of the mass function ($m( \mathbf{E}, \mathbf{h})$) enforce the zero conditions, 
\begin{eqnarray}
m( \mathbf{h}, \mathbf{E}) =0, \quad  \mathbf{E} \parallel \hat{\mathbf{b}} \,\,\,{\rm or}\,\,\,  \mathbf{E} \parallel  \hat{\mathbf{c}},  \label{equation1}
\end{eqnarray}
with magnetic fields along the bond directions, $\mathbf{h} \parallel \hat{\mathbf{b}}$. 
The zero conditions guarantees  the existence of the gapless Majorana excitations. 
Second, the symmetry properties of the chemical potential function give the zero condtion, 
\begin{eqnarray}
\mu( \mathbf{h}, \mathbf{E}) =0, \quad  \mathbf{E} \parallel \hat{\mathbf{b}}, \label{equation2}
\end{eqnarray}
with magnetic fields along the bond directions, $\mathbf{h} \parallel \hat{\mathbf{b}}$. 
On the other hand, $\mu( \mathbf{h}, \mathbf{E})$ is not generically zero for $\mathbf{E} \parallel \hat{\mathbf{c}}$, which indicates that  the Majorana Fermi surfaces may appear near $\mathbf{E} \parallel \hat{\mathbf{c}}$ because $|\mu( \mathbf{h}, \mathbf{E})|$ is generically bigger than $|m( \mathbf{h}, \mathbf{E})|$.

\begin{figure}[tb]
\includegraphics[width=0.48\textwidth]{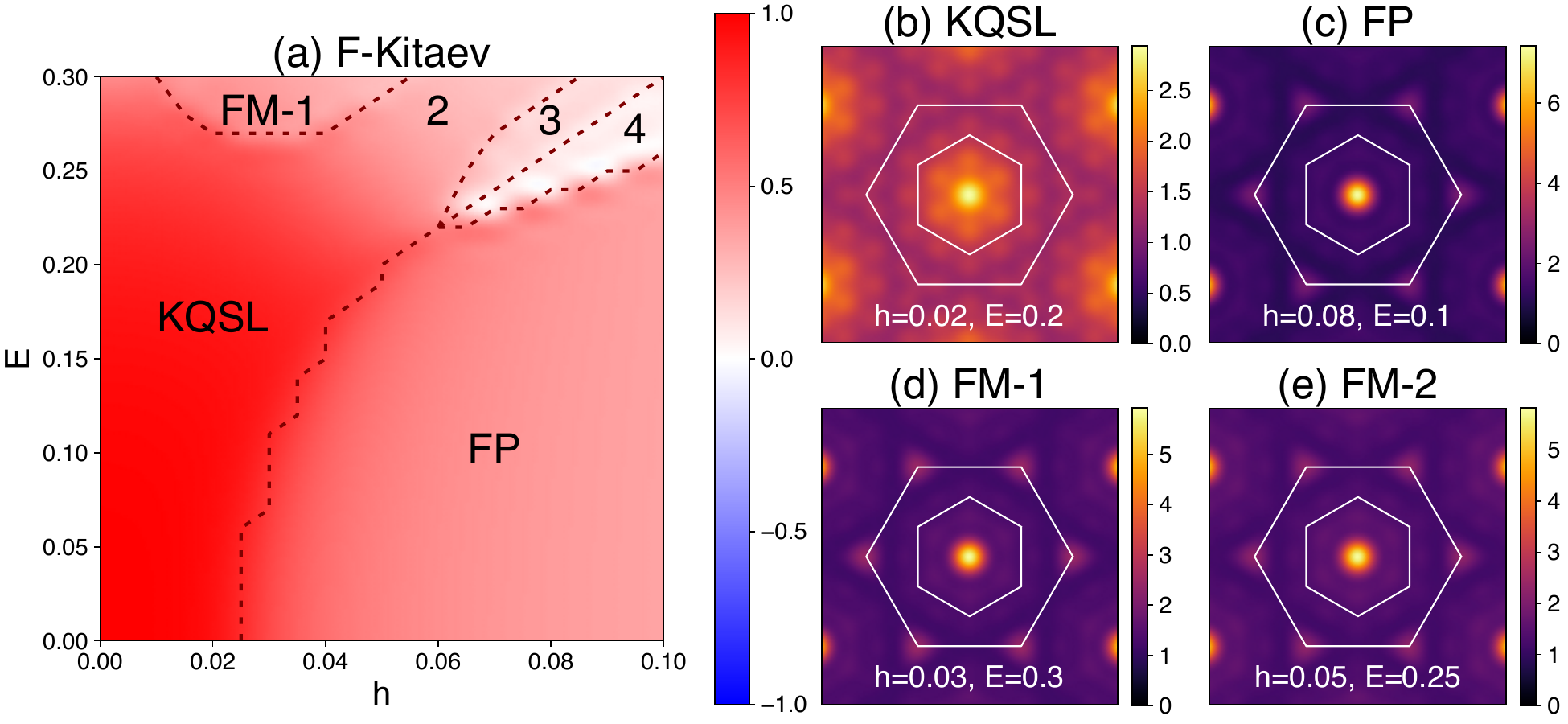}
  \caption{
Phase diagram of the ferromagnetic Kitaev system ($K = -1$). (a) The phase diagram in the plane of $h$ and $E$. FM-1,2,3,4: ferromagnetically ordered phases. FP: field polarized phase. The color code means the $\mathbb{Z}_2$ flux expectation value $\langle \hat{W}_p \rangle$, and the dashed lines indicate the phase boundaries determined by the ground state energy second derivative $-\partial^2 E_{gs}/\partial \xi^2~(\xi = h,E)$. (b-e) Spin structure factors for different phases. In each plot, the two hexagons denote the first and second Brillouin zones in momentum space. In all the results, the magnetic field and electric field are both aligned along the $c$-axis (${\bf h}, {\bf E}\parallel{\bf c}$). 
}
 \label{fig:2}
\end{figure}

\textit{Exact diagonalization.} 
We further solve the model Hamiltonian on a 24-site cluster with the periodic boundary condition by using exact diagonalization.
We determine the phase diagram for ferromagnetic Kitaev interaction (see SM for antiferromagnetic Kitaev interaction) by computing (i) the ground state energy second derivatives, $-\partial^2 E_{\rm gs}/\partial \xi^2~(\xi=h,E)$, (ii) the $\mathbb{Z}_2$ flux, $\langle \hat{W}_p \rangle$, and (iii) the spin structure factor, $S({\bf q})=\frac{1}{N}\sum_{i,j}\langle {\bf S}_i\cdot{\bf S}_j\rangle e^{i{\bf q}\cdot{({\bf r}_i-{\bf r}_j})}$, as illustrated in Figure~\ref{fig:2}. The electric and magnetic fields are along the $c$-axis (${\bf h}, {\bf E}\parallel{\bf c}$) for illustration.

A few remarks are as follows. 
First, we find that the KQSL phase with a ferromagnetic Kitaev interaction is more stable under an electric field than a magnetic field while the KQSL with an antiferromagnetic Kitaev interaction is more stable under a magnetic field as shown in SM. 
Second, the upper critical electric field is $E^{u}_c \approx0.27$ without a magnetic field, critical magnetic field is $h_c \approx0.03$ without a electric field and increases turning on an electric field. 
The increase indicates that electric and magnetic fields are synergetic to stabilize KQSLs.  
Third, we also find nearby phases marked by dashed lines. 
The ferromagnetic phases (FM-1,2,3,4) show spin moment canting within the plane and out of the plane due to the electric and magnetic fields. 
Although they are distinguished by the ground state energy second derivatives $-\partial^2 E_{\rm gs}/\partial \xi^2~(\xi=h,E)$, we do not find any qualitative difference among the FM phases [Fig.~\ref{fig:2}(d,e)]. 
Fourth, the introduction of non-Kitaev interactions such as the Heisenberg interaction modifies phase diagrams quantitatively but not qualitatively. 

Based on the results of exact diagonalization, we conclude that the stability of the KQSLs under electric and magnetic fields is guaranteed though their critical field values depend on details of the microscopic Hamiltonian. Then, the symmetry properties of Majorana fermions become very useful for stable KQSLs. Namely, the two zero conditions, Eqns \eqref{equation1} and \eqref{equation2}, are solely determined by the symmetry properties of $\mathbb{G}_{b,c}$, indicating that the zero conditions even work beyond the pure Kitaev model. It is straightforward to show that non-Kitaev interaction terms induce effective interaction between gapless Majorana fermions which are known to be irrelevant in the sense of renormalization group analysis,  in addition to trivial renormalization of velocity.  
From now on, unless stated otherwise, we utilize the symmetry properties and our results hold beyond the pure Kitaev model.

\textit{Two types of TQPTs.} 
In KQSLs, we uncover the two types of TQPTs. The first one is conventional in the sense that topological phases with $\nu=\pm 1$ are generically connected through a quantum critical point,  named type-I TQPT. In other words, gapless Majorana fermions appear only at quantum critical points. 
Not only the zero conditions ($m(\mathbf{h}, \mathbf{E})=\mu(\mathbf{h}, \mathbf{E})=0$) but also the exclusion of Majorana Fermi surfaces are necessary to find such quantum critical points. 
The former is satisfied by $\mathbf{E} \parallel \hat{\mathbf{b}}$ and the latter is fulfilled by $|m( \mathbf{h}, \mathbf{E})| > |\mu( \mathbf{h}, \mathbf{E})|$ near  $\mathbf{E} \parallel \hat{\mathbf{b}}$.
Then, we obtain the condition of the type-I TQPTs, 
%
 \begin{eqnarray}
 \mathbf{E} \parallel \hat{\mathbf{b}}, \quad |\mathbf{E}| > E^{l}_c,  \quad {\rm (Type \,I)}
\end{eqnarray}
where the lower critical electric field ($E^{l}_c$) is to exclude Majorana Fermi surfaces. Its value is determined by microscopic information. For example, the pure Kitaev model gives $E^{l}_c = (c_{\mu} h^2)/(2 c_m \Delta_f)$, and the critical points are illustrated in the dashed blue line in in Figure \ref{fig1}(b).
The second one is unconventional in the sense that topological phases with $\nu=\pm 1$ are generically connected through quantum critical states with Majorana Fermi surfaces, named Type-II TQPT. The transition lines are determined by the condition,
 \begin{eqnarray}
|m( \mathbf{h}, \mathbf{E})| = |\mu( \mathbf{h}, \mathbf{E})|>0,  \quad {\rm (Type \,II)}
\end{eqnarray}
as illustrated in the dotted black line in Fig. 1(b). 
We note that the type-II TQPT completely disappear with the fine-tuned condition, $c_{\mu}=0$, for the pure Kitaev model. 
In other words, the presence of $\mu( \mathbf{h}, \mathbf{E})$ is essential to the presence of type-II TQPTs, and both the electric and magnetic fields are necessary, as pointed out previously \cite{Chari2021.4}.

\begin{figure}[t]
\includegraphics[width=0.48\textwidth]{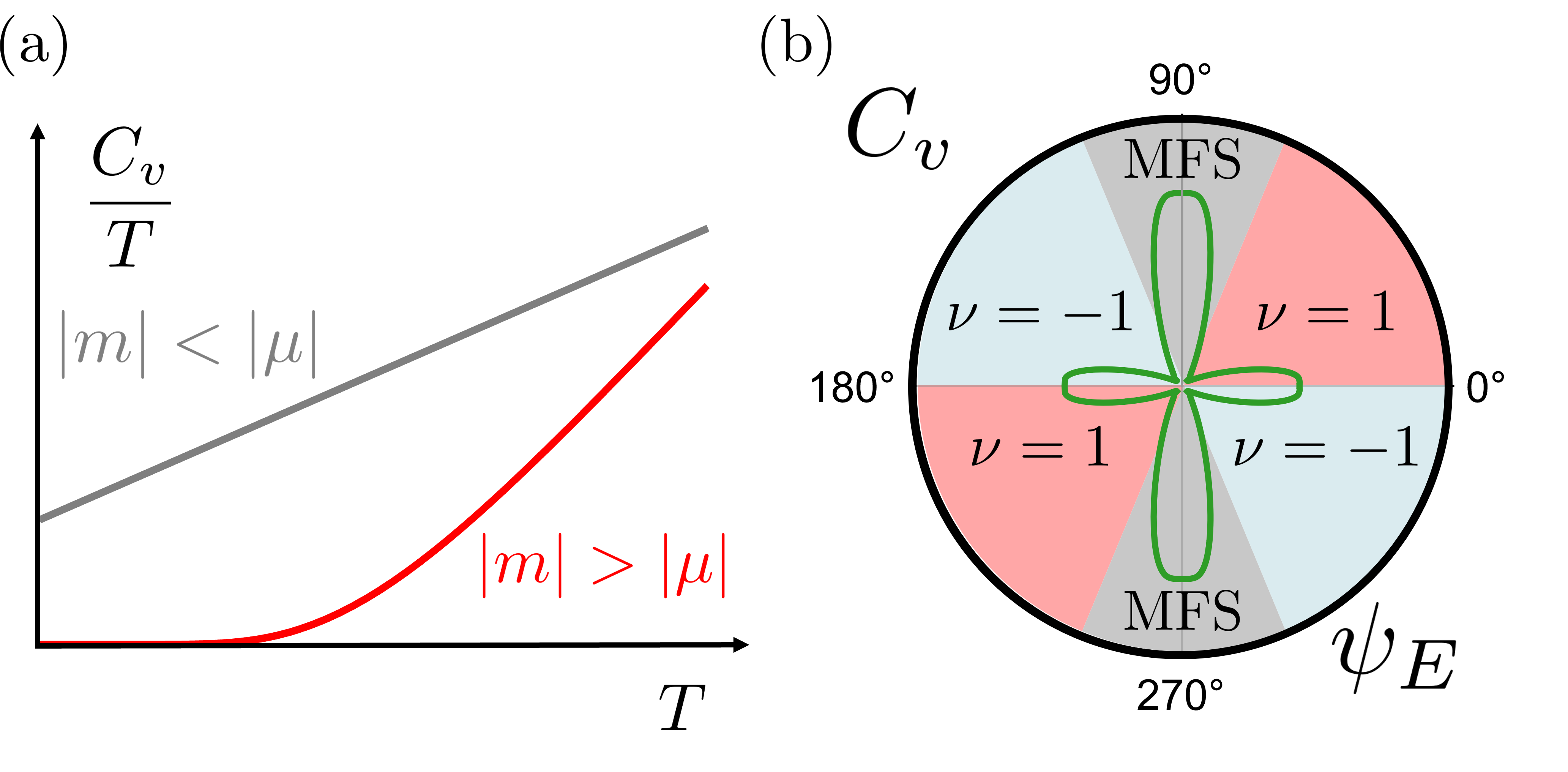}
  \caption{  Schematic low temperature specific heat ($C_v$)with magnetic fields and electric fields. (a)  Temperature $(T)$ dependence of  $\frac{C_{v}}{T}$ with fixed electric and magnetic fields. (b) Angle dependence ($\psi_E$) of specific heat for $E >E^{l}_c$  with a fixed temperature. }
\label{fig5}
\end{figure}

{\it Discussion and conclusion: }
We propose that electric-field-driven TQPTs may be utilized to identify KQSLs.
Varying with the amplitude of electric fields, TQPTs between critical states and bulk energy-gapped states generically appear. 
With $\mathbf{h} \parallel \hat{\mathbf{b}}$, a small electric field ($E < E^{l}_c$) cannot introduce an energy gap of Majorana fermions while a large electric field ($E^{l}_c<E<E^{u}_c $) induces a topological phase with a well-defined bulk energy gap. Such phase transitions may be readily observable in specific heat experiments, which carry low-energy excitations, as illustrated in Figure~\ref{fig5}(a). 
Furthermore, the rotation of an electric field is a natural way to observe the two types of TQPTs for $ E^{l}_c < E < E^{u}_c$. 
We note that the type-II TQPTs around $\mathbf{E} \parallel \hat{\mathbf{c}}$ are are natural outcomes of the zero conditions, Eqns \eqref{equation1} and \eqref{equation2}, which  give a non-zero value of specific heat over temperature ($C_v/ T$) in the zero temperature limit. Thus, the field angle dependence specific heat naturally has the two-fold symmetric behavior as illustrated in Figure~\ref{fig5}(b). 
Such characteristics are in drastic contrast to other paramagnetic phases including partially polarized phases whose ground states are adiabatically connected to a simple product state without quantum entanglement \cite{Chern2021.4}. 

We further discuss energy scales of electric-field-driven TQPTs. Setting the Kitaev interaction term as unity, it is well known that the flux energy gap is $\Delta_f \sim 0.07$ \cite{Kitaev2005.10} and the KQSL is stable below $h_c \sim 0.03$, which gives the lower critical electric field ($E^{l}_c \sim 0.006$) assuming that $c_{\mu}$ and $c_{m}$ are same order of magnitude.   
For real materials \cite{Winter2017.11,Bolens2018.4,Chari2021.4}, the amplitude electric field is estimated as  $E \sim 4\times 10^{-3}\text{meV}\sim 0.01\Delta_f $ for the strength of electric field $10^6$V/m. 
Though our estimation needs to be scrutinized for real candidate materials, we expect that electric-field-driven TQPTs may be observable in experiments. 
%

%

In conclusion, we investigate the electric-field-driven TQPTs in KQSLs. 
In sharp contrast to the common belief that an insulator is inert under weak electric fields due to charge energy gaps, KQSLs may host significant effects with small electric fields because of non-trivial symmetry properties of Majorana fermions of KQSLs. 
We find TQPTs between critical states and bulk energy-gapped states varying with the amplitude of electric fields. Also, by rotating an electric field, we find the possibility of the two types of TQPTs between the phases with opposite topological invariants.  
Such TQPTs are associated with characteristic structures of gapless excitations, and thus we propose intriguing specific heat signatures in candidate materials of KQSLs such as $\alpha$-RuCl$_3$.

{\it Acknowledgements : } 
We thank  Ara Go and Takasada Shibauchi for earlier collaboration and invaluable discussion about experimental setups. 
P.N. and E.-G.M. were supported by the National Research Foundation of Korea funded by the Ministry of Science and ICT (No. 2021R1A2C4001847, No. 2022M3H4A1A04074153, No. 2023M3K5A1094813) and National Measurement Standard Services and Technical Services for SME funded by Korea Research Institute of Standards and Science (KRISS – 2022 – GP2022-0014). K.H. was supported by Individual Grant (No. PG071403) of Korea Institute for Advanced Study (KIAS) where computations were performed on clusters at the Center for Advanced Computation.

\bibliographystyle{apsrev}
\bibliography{references.bib}
 
\clearpage
\onecolumngrid
\begin{center}
\textbf{\large Supplementary Material for \\``Manipulating Topological Quantum Phase Transitions \\ of Kitaev's Quantum Spin Liquids with Electric Fields''} 
\end{center}

\newcommand{\slantedparallel}{\mathbin{\!/\mkern-5mu/\!}}
\renewcommand{\thefigure}{S\arabic{figure}}
\renewcommand{\thetable}{S\Roman{table}} 
\renewcommand{\theequation}{S\arabic{equation}} 

\setcounter{equation}{0}
\setcounter{figure}{0}
\setcounter{table}{0}
\setcounter{page}{1}

\section{1. Polarization Operator}
In this section, we provide more information on the polarization operator. Following the previous works  \cite{Miyahara2016.1,Bolens2018.4,Bolens2018.9,Chari2021.4}, we consider only nearest neighbors two-spin terms. Since the polarization oerator is even under the time-reversal transformation and odd under the space-inversion transformation, the polarization operator ($\mathbf{P}$) should be, 
\begin{align*}
P^{\mu}=\sum_{\langle i,j \rangle_{\gamma}}\mathbf{p}^{\mu}_{\gamma} \cdot (\mathbf{S}_i \times \mathbf{S}_j)
\end{align*}
where $\mu = x, y, z$ axis, $\gamma = x, y, z$-bond. and $\mathbf{p}^{\mu}_{\gamma}$ is vector in three dimensions. Therefore, there are 27 parameters related to the polarization operator. From $D_3$ symmetry, these 27 parameters are reduced to 5 parameters ($c_1 \sim c_5$),
\begin{align*}
&\mathbf{p}^{\alpha}_{\alpha}=c_1 \hat{\alpha}+c_2( \hat{\beta}+ \hat{\gamma})\\
&\mathbf{p}^{\alpha}_{\beta}=c_3 \hat{\alpha}+c_4 \hat{\beta}+c_5 \hat{\gamma}\\
\end{align*}
where $(\alpha, \beta, \gamma) $are any permutations of $(x, y, z)$. The Hamiltonian for electric field ($H_E$) is
\begin{align*}
&H_E=-\mathbf{E}\cdot \mathbf{P}=-\sum_{\langle i,j \rangle_{\gamma}} \left( \sum_{\mu} E_{\mu}\mathbf{p}^{\mu}_{\gamma}\right) \cdot (\mathbf{S}_i \times \mathbf{S}_j)=-\sum_{\langle i,j \rangle_{\gamma}} \mathbf{d}_{\gamma} \cdot (\mathbf{S}_i \times \mathbf{S}_j) 
\end{align*}
where $\mathbf{d}_{\gamma}=\left( \sum_{\mu} E_{\mu}\mathbf{p}^{\mu}_{\gamma}\right) $,
\begin{align*}
&d^{\alpha}_{\alpha}=c_1E_{\alpha}+c_4(E_{\beta}+E_{\gamma})\\
&d^{\alpha}_{\beta}=c_3E_{\alpha}+c_2E_{\beta}+c_5E_{\gamma}
\end{align*}
where $ (\alpha,\beta, \gamma)$ are any permutations of $(x, y, z)$. For simple case ($c=c_1=c_3$, $c_2=c_4=c_5=0$), we can simplify the Hamiltonian $(H_E)$.
\begin{align*}
&H_E=-(c\mathbf{E}) \cdot \sum_{\langle i,j \rangle_{\gamma}}  (\mathbf{S}_i \times \mathbf{S}_j) 
\end{align*}
One of microscopic estiamtion of $c$ is $c \sim 3.5 \times 10^{-9} \frac{\text{meV}}{\text{(V/m)}}$ \cite{Bolens2018.4,Bolens2018.9,Chari2021.4}. We absorb the $c$  into the $\mathbf{E}$ in the main text for the sake of convenience,
\begin{align*}
H_E=-\mathbf{E} \cdot \sum_{\langle i,j \rangle_{\gamma}}  (\mathbf{S}_i \times \mathbf{S}_j) 
\end{align*}
\section{2. Majorana operator representation}
In this section, we provide more inforamtion on Majorana operator representation. 
%
We can express the isotropic Kitaev Hamiltonain ($H_{K}$) as Majorana oerators ($\mathbf{b}_j,c_j$) from a Majorana representation $\boldsymbol{\sigma}_j=i\mathbf{b}_jc_j$ with constraint $D_j\equiv b^{x}_jb^{y}_jb^{z}_jc_j=1$ for each site $j$ \cite{Kitaev2005.10},
\begin{align*}
&H_{K}=\frac{1}{2}\sum_{\mathbf{k}} 
\Psi_{\mathbf{k}}^{\dagger}
\begin{pmatrix}
    0 & F(k)\\
F^{*}(k)&0
\end{pmatrix}
\Psi_{\mathbf{k}},\quad  F(k)=-\frac{iK}{2}\left(e^{i\mathbf{k}\cdot \mathbf{n}_1}+e^{i\mathbf{k}\cdot \mathbf{n}_2}+1\right)
\end{align*}
where $\Psi_{\mathbf{k}}=(c_{\mathbf{k},A},c_{\mathbf{k},B})^{T}$,  and $c_{\mathbf{r},A(B)}=\sqrt{\frac{2}{N}}\sum_{\mathbf{k}}e^{i\mathbf{k}\cdot \mathbf{r}}c_{\mathbf{k},A(B)} $. $A, B (\mathbf{n}_1,\mathbf{n}_2)$ are sublattice index (basis vectors for lattice) \cite{Kitaev2005.10}.

When the magnetic field and electric field are turned on, one of methods to handle is quai-degenerate perturbation. 
The main idea is that we will substitute any resolvents in the perturbation with a single energy scale, the relevant flux gap ($\Delta_f\sim 0.07|K|$). Then, we can construct effective Hamiltonian ($H_{\text{eff}}$) in the flux-free sector when additional terms are handled perturbatively,

\begin{align*}
&H=K \sum_{\langle i,j\rangle_{\gamma}}S^{\gamma}_iS^{\gamma}_j -\sum_{j}\mathbf{h}\cdot \mathbf{S}_j -\sum_{\langle i,j\rangle_{\gamma}} \mathbf{d}_{\gamma}\cdot (\mathbf{S}_i \times \mathbf{S}_j),\\
&H_{\text{eff}}=\frac{1}{2}\sum_{\mathbf{k}} 
\Psi_{\mathbf{k}}^{\dagger}
\left(
\sum_{a=0,1,2, 3} \epsilon_a(\mathbf{k}, \mathbf{h}, \mathbf{E}) \tau^a 
\right)\Psi_{\mathbf{k}},
\end{align*}
where $\tau^{1,2,3}$ are usual Pauli matrices and $\tau^{0}$ is the identity matrix.
Because original Kitaev Hamiltonian ($H_K=K \sum_{\langle i,j\rangle_{\gamma}}S^{\gamma}_iS^{\gamma}_j$) in the system, the off-diagonal terms are,
\begin{align*}
&\epsilon_1(\mathbf{k}, \mathbf{h}, \mathbf{E})\tau^1+\epsilon_2(\mathbf{k}, \mathbf{h}, \mathbf{E})\tau^2 \approx 
\begin{pmatrix}
    0 & F(k)\\
F^{*}(k)&0
\end{pmatrix}
\end{align*}
and, we get the diagonal terms up to third order,
\begin{align*}
\epsilon_0(\mathbf{k}, \mathbf{h}, \mathbf{E})&=-\frac{1}{\Delta_f}((h_xd^{x}_y-h_yd^{y}_x)\sin{(\mathbf{k}\cdot(\mathbf{n}_1-\mathbf{n}_2))}+(h_yd^{y}_z-h_zd^{z}_y)\sin{(\mathbf{k}\cdot \mathbf{n}_2)}+(h_zd^{z}_x-h_xd^{x}_z)\sin{(-\mathbf{k}\cdot\mathbf{n}_1)})\\
\epsilon_3(\mathbf{k}, \mathbf{h}, \mathbf{E})&=G_1(\mathbf{h},\mathbf{E})\sin{(\mathbf{k}\cdot(\mathbf{n}_1-\mathbf{n}_2))}+G_1(C_3\mathbf{h},C_3\mathbf{E})\sin{(\mathbf{k}\cdot \mathbf{n}_2)}+G_1(C^2_3\mathbf{h},C^2_3\mathbf{E})\sin{(-\mathbf{k}\cdot \mathbf{n}_2)}\\
&=G_2(\mathbf{h},\mathbf{E})\sin{(2\mathbf{k}\cdot(\mathbf{n}_1-\mathbf{n}_2))}+G_2(C_3\mathbf{h},C_3\mathbf{E})\sin{(2\mathbf{k}\cdot \mathbf{n}_2)}+G_2(C^2_3\mathbf{h},C^2_3\mathbf{E})\sin{(-2\mathbf{k}\cdot \mathbf{n}_2)}\\
&=G_3(\mathbf{h},\mathbf{E})\sin{(\mathbf{k}\cdot(2\mathbf{n}_2-\mathbf{n}_1))}+G_3(C_3\mathbf{h},C_3\mathbf{E})\sin{(\mathbf{k}\cdot (2\mathbf{n}_2-\mathbf{n}_1))}+G_3(C^2_3\mathbf{h},C^2_3\mathbf{E})\sin{(-\mathbf{k}\cdot (\mathbf{n}_2+\mathbf{n}_1))},\\
\end{align*}
\begin{align*}
G_1(\mathbf{h},\mathbf{E})&=\frac{1}{4\Delta_f^2}(-12h_xh_yh_z+h_x(3d^{y}_{y}d^{z}_{y}-d^{y}_{x}d^{z}_{y}+d^{z}_{x}d^{y}_{y}-3d^{y}_x(d^{z}_{x}+d^{z}_{z}))\\
&+h_y(3d^{z}_{x}d^{x}_{x}-d^{z}_{x}d^{x}_{y}+d^{x}_{x}d^{z}_{y}-3d^{x}_y(d^{z}_{y}+d^{z}_{z}))+h_z(3d^{x}_y(d^{y}_{y}+d^{y}_{z})+3d^{y}_x(d^{x}_{x}+d^{x}_{z})+3(d^{x}_xd^{y}_z+d^{x}_zd^{y}_z+d^{x}_zd^{y}_y)+3d^{x}_yd^{y}_x)\\
G_2(\mathbf{h},\mathbf{E})&=\frac{3h_zd^{y}_xd^{x}_y}{4\Delta_f^2}\\
G_3(\mathbf{h},\mathbf{E})&=\frac{3}{4\Delta_f^2}((h_zd^{y}_x(d^{x}_z+d^{x}_x)-h_yd^{z}_x(d^{x}_y+d^{x}_x))\\
\end{align*}
where $C_3$ is the  three-fold rotation operator such that $C_3(h_{x,y,z},E_{x,y,z})=(h_{y,z,x},E_{y,z,x})$.

\section{3. Symmetry Properties}

\begin{table}[tb]
\centering 
\begin{tabular}{>{\centering}p{6cm}||>{\centering}p{1.1cm}|>{\centering}p{1.1cm}|>{\centering}p{2.2cm}|>{\centering\arraybackslash} p{2.2cm}}
\hline \hline
Majorana Operator & $\mathcal{T}$ & $\mathcal{P}$ & $C_{2x},C_{2y},C_{2z}$ & $C_3,C^2_3$ \\ 
\hline \hline 
 $\Psi_{K}^{\dagger} \tau^{0} \Psi_{K}$ & odd &odd & odd & even \\

 $\Psi_{K}^{\dagger} \tau^{1} \Psi_{K}$  & even & even & odd & even \\ 

 $\Psi_{K}^{\dagger} \tau^{2} \Psi_{K}$ & even & even & even & even  \\

 $\Psi_{K}^{\dagger} \tau^{3} \Psi_{K}$ & odd & even & odd &even  \\

\hline \hline
$m( \mathbf{h}, \mathbf{E})$ & $\mathcal{T}$ & $\mathcal{P}$ & $C_{2x},C_{2y},C_{2z}$ & $C_3,C^2_3$ \\ 
\hline \hline 
$h_c$, $h_c^3$, $h_cE_c^2$, $h_a(h_a^2-3h_b^2)$ & odd & even & odd & even \\

$h_c(h_a^2+h_b^2)$, $h_c(E_a^2+E_b^2)$  & odd & even & odd & even \\ 

$h_aE_a^2-h_aE_b^2-2h_bE_aE_b$ & odd & even & odd & even  \\

 $E_c(E_ah_a+E_bh_b)$  & odd & even & odd &even  \\
\hline\hline
%
$\mu( \mathbf{h}, \mathbf{E})$& $\mathcal{T}$ & $\mathcal{P}$ & $C_{2x},C_{2y},C_{2z}$ & $C_3,C^2_3$ \\ 
\hline  \hline
$(h_a E_b - h_b E_a)$, $E_ch_b(h_b^2-3h_a^2)$ & odd & odd & odd & even \\
$E^2_c(h_a E_b - h_b E_a)$,$h^2_c(h_a E_b - h_b E_a)$ & odd & odd & odd & even  \\

$E_c(h_bE_b^2-h_bE_a^2-2h_aE_aE_b)$ & odd & odd & odd & even  \\

 $-h_a(E_b^3+E_bE_a^2)+h_b(E_a^3+E_aE_b^2)$ & odd & odd & odd & even  \\

 $h_c(E_bh_b^2-E_bh_a^2-2E_ah_ah_b)$& odd & odd & odd & even  \\

$-E_a(h_b^3+h_bh_a^2)+E_b(h_a^3+h_ah_b^2)$ & odd & odd & odd & even  \\

 $h_cE_b(E_b^2-3E_a^2)$& odd & odd & odd & even  \\

\hline
\hline

\end{tabular}\\ 

\caption{Irreducible representions for bilinear Majorana operators at the $K$ point, mass function and chemical potential function. We use the  $\hat{\mathbf{a}}=\frac{1}{\sqrt{6}}(\hat{\mathbf{x}}+\hat{\mathbf{y}}-2\hat{\mathbf{z}})$, $\hat{\mathbf{b}}=\frac{1}{\sqrt{2}}(-\hat{\mathbf{x}}+\hat{\mathbf{y}})$, $\hat{\mathbf{c}}=\frac{1}{\sqrt{3}}(\hat{\mathbf{x}}+\hat{\mathbf{y}}+\hat{\mathbf{z}})$, and $\Psi_{K}=(c_{\mathbf{K},A},c_{\mathbf{K},B})^{T}$.}
 \label{supple_table1}
\end{table}

In this section, we discuss properties of symmetry properties of Majorana operators and fields to get the general form of mass/chemical potential functions ($m(\mathbf{h},\mathbf{E})$/$\mu(\mathbf{h},\mathbf{E})$) in the main text.
We are interested in  the symmetry group, $ \mathbb{D}_3\otimes \mathbb{P}\otimes\mathbb{T} $, where $\mathbb{P}, \mathbb{T}$ are spacial inversion, time reversal, and $\mathbb{D}_3$ is the group consisting of $C_3$ rotation around out of plane direction and two-fold rotations around each bond directions ($C_{2x},C_{2y},C_{2z}$),
\begin{align*}
&\mathbb{P}=\{I,\mathcal{P}\},\quad \mathbb{T}=\{I,\mathcal{T}\},\quad \mathbb{D}_3=\{I,C_3,C_3^2,C_{2x},C_{2y},C_{2z}\},
\end{align*}
where $I$ is the identity transform. The character tables for the $\mathbb{P}\otimes\mathbb{T} $ group and $\mathbb{D}_3$ group are, 
\begin{center}
\begin{tabular}{>{\centering}p{1cm}||>{\centering}p{1cm}>{\centering}p{1cm}>{\centering}p{1cm}>{\centering}p{1cm}||>{\centering}p{1cm}||>{\centering}p{1cm}>{\centering}m{1cm}>{\centering\arraybackslash}p{2cm}}
\noalign{\smallskip}\noalign{\smallskip}\hline\hline
$\mathbb{P}\otimes\mathbb{T} $& I & $\mathcal{P}$ & $\mathcal{T}$& $\mathcal{PT}$&$\mathbb{D}_3$& I & $3C_2$ & $2C_3$ \\
\hline
\hline
$\mathscr{B}_{I}$ & 1 & 1 & 1& 1&$\mathscr{A}_1$ & 1 & 1 & 1 \\

$\mathscr{B}_{\mathcal{T}}$& $1$ & $1$ &$-1$& $-1$&$\mathscr{A}_{2}$& $1$ & $-1$ &$1$\\
$\mathscr{B}_{\mathcal{P}}$  & $1$& $-1$ & $1$ & $-1$&\multirow{2}{*}{$\mathscr{E}$}  & \multirow{2}{*}{$2$}& \multirow{2}{*}{$0$} & \multirow{2}{*}{$-1$}\\
$\mathscr{B}_{\mathcal{PT}}$  & $1$ & $-1$ & $-1$& $1$ &&&&\\
\hline
\hline
\end{tabular}
\end{center}
\begin{center}
The character tables for the $\mathbb{P}\otimes\mathbb{T}$, and $\mathbb{D}_3 $ 
\end{center}

We need to know how Majorana operators transform with respect to the symmetries we considered
Because we
consider flux-free sector, and gauge choice ($ib_jb_k = 1$, $j \in A$ sublattice, $k \in B$ sublattice), Majorana operators change
effectively \cite{Kitaev2005.10},
\begin{align*}
g&: \quad c_{\mathbf{r},A}\to c_{g(\mathbf{r}),A},\; c_{\mathbf{r},B}\to c_{g(\mathbf{r}),B},\;\text{where } g\in \mathbb{D}_3, \\
\mathcal{T}&:\quad c_{\mathbf{r},A}\to c_{\mathbf{r},A},\; c_{\mathbf{r},B}\to -c_{\mathbf{r},B},\\
\mathcal{P}&:\quad c_{\mathbf{r},A}\to c_{-\mathbf{r},B},\; c_{\mathbf{r},B}\to -c_{-\mathbf{r},A},
\end{align*}
As for the $k$-space ($c_{\mathbf{k},A(B)}\propto \sum_{\mathbf{r}}e^{-i\mathbf{r}\cdot \mathbf{k}} c_{\mathbf{r},A(B)}$),
\begin{align*}
g&: \quad c_{\mathbf{k},A}\to c_{g(\mathbf{k}),A},\; c_{\mathbf{k},B}\to c_{g(\mathbf{k}),B},\;\text{where } g\in \mathbb{D}_3, \\
\mathcal{T}&:\quad c_{\mathbf{k},A}\to c_{-\mathbf{k},A},\; c_{\mathbf{k},B}\to -c_{-\mathbf{k},B},\\
\mathcal{P}&:\quad c_{\mathbf{k},A}\to c_{-\mathbf{k},B},\; c_{\mathbf{k},B}\to -c_{-\mathbf{k},A},
\end{align*}
We can check $ \mathbb{T} \bigotimes \mathbb{P}$ act on singe Majorana operator projectively  \cite{Kitaev2005.10},
\begin{align*}
\mathcal{PT}(c_{\mathbf{k},A})= c_{\mathbf{k},B}\neq  -c_{\mathbf{k},B}=\mathcal{T P}(c_{\mathbf{k},A}).
\end{align*}

We get general form of mass/chemical potential functions ($m(\mathbf{h},\mathbf{E})$/$\mu(\mathbf{h},\mathbf{E})$) from three steps.

\begin{itemize}
\item[Step 1.] We find how bilinear Majorana operators at the $K$ point behave under $ \mathbb{D}_3\otimes \mathbb{P}\otimes\mathbb{T}$ as shown in the Table \ref{supple_table1}. Note that $\Psi_{K}^{\dagger} \tau^{0} \Psi_{K}$ ($\Psi_{K}^{\dagger} \tau^{3} \Psi_{K}$) is $\mathscr{A}_{2}\otimes \mathscr{B}_{\mathcal{PT}}$ ($\mathscr{A}_{2}\otimes \mathscr{B}_{\mathcal{T}}$) irreducible representation. It implies that $\mu(\mathbf{h},\mathbf{E})$ $(m(\mathbf{h},\mathbf{E}))$ is also $\mathscr{A}_{2}\otimes \mathscr{B}_{\mathcal{PT}}$ ($\mathscr{A}_{2}\otimes \mathscr{B}_{\mathcal{T}}$) irreducible representation.
\item[Step 2.] We give all irreducible representations of electric field ($\mathbf{E}$), magnetic field ($\mathbf{h}$) for $\mathscr{A}_{2}\otimes \mathscr{B}_{\mathcal{PT}}$ and $\mathscr{A}_{2}\otimes \mathscr{B}_{\mathcal{T}}$ up to forth order as shown the Table \ref{supple_table1}. 
\item[Step 3.]
From the representations, we write down general form of $m(\mathbf{h},\mathbf{E}),\mu(\mathbf{h},\mathbf{E})$,
\begin{align*}
m(\mathbf{h},\mathbf{E})=&k_1h_c+k_2h_a(h_a^2-3h_b^2)+k_3h_c^3+k_4h_c(h_a^2+h_b^2)+k_5h_c(E_a^2+E_b^2)+k_6(h_aE_a^2-h_aE_b^2-2h_b E_aE_b)\\
&+k_7h_cE_c^2+k_8E_c(E_ah_a+E_bh_b),\\
\mu(\mathbf{h},\mathbf{E})=&k_9(E_ah_b-E_bh_a)+k_{10}h_cE_b(E_b^2-3E_a^2)+k_{11}  E_c(h_bE_b^2-h_bE_a^2-2h_aE_aE_b)\\
&+k_{12}(-h_a(E_b^3+E_bE_a^2)+h_b(E_a^3+E_aE_b^2))+k_{13}E_ch_b(h_b^2-3h_a^2)+k_{14}h_c(E_bh_b^2-E_bh_a^2-2E_ah_ah_b)\\
&+k_{15}(-E_a(h_b^3+h_bh_a^2)+E_b(h_a^3+h_ah_b^2))+k_{16}E_c^2(E_ah_b-E_bh_a)+k_{17}h_c^2(E_ah_b-E_bh_a)
\end{align*}
where $k_1$ to $k_{17}$ are some real constants.
\end{itemize}

\section{4. Additional results of exact diagonalization}

\begin{figure*}[tb]
\includegraphics[width=\linewidth]{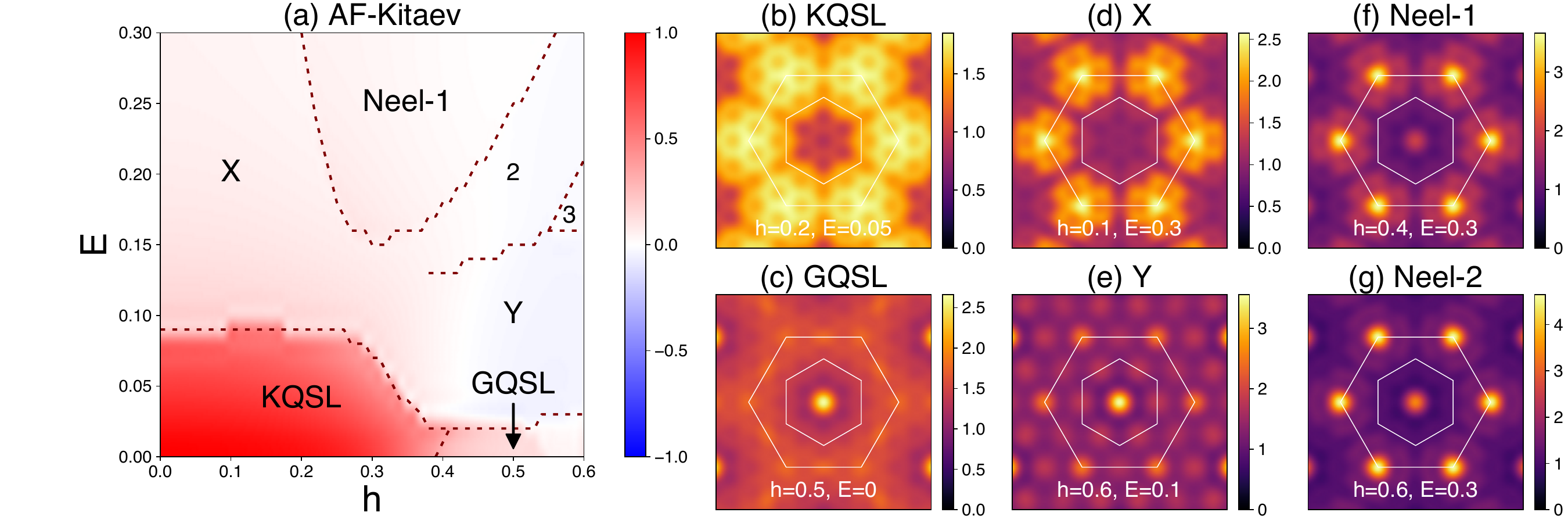}
\caption{Antiferromagnetic Kitaev system ($K=1$). (a) The phase diagram in the plane of the magnetic and electric fields along the $c$-axis (${\bf h}, {\bf E} \parallel {\bf c}$). The color represents the $\mathbb{Z}_2$ flux expectation value $\langle \hat{W} \rangle$, and the dashed lines indicate the phase boundaries determined by the ground state energy second derivative $-\partial^2 E_{\rm gs}/\partial \xi^2 ~(\xi=h,E)$. (b-g) Spin structure factors for several selected points. The two hexagons denote the first and second Brillouin zones in momentum space.}
\label{fig:AF-Kitaev}
\end{figure*}

We provide results of exact diagonalization for the system with antiferromagnetic Kitaev interaction ($K=1$). Figure~\ref{fig:AF-Kitaev} shows the phase diagram and the spin structure factor. 
In this case, a gapless quantum spin liquid (GQSL) exists in addition to the Kitaev quantum spin liquid (KQSL) at low electric fields~\cite{Hickey2019.1}. At high electric and magnetic fields, we find antiferromagnetically ordered Neel phases (Neel-1,2,3). Similar to the FM phases in the ferromagnetic Kitaev system, these Neel phases show spin moment canting due to the electric and magnetic fields, and also they look qualitatively same. Interestingly, some intermediate phases (X and Y) arise between the Neel phases and the KQSL and GQSL phases [Fig.~\ref{fig:AF-Kitaev}(a)]. The X and Y phases show broad features in the spin structure factor rather than sharp peaks, suggesting possibilities of disordered phases. See the spin structure factors of the X and Y phases in Fig.~\ref{fig:AF-Kitaev}(d,e).

\end{document}